# Electrically pumped *h*-BN single-photon emission in van der Waals heterostructure

*Mihyang Yu,[1] Jeonghan Lee,[1] Kenji Watanabe,[2] Takashi Taniguchi,[3] and Jieun Lee[1*]*

[1]Department of Physics and Astronomy, Seoul National University, Seoul, 08826, Korea

[2]Research Center for Electronic and Optical Materials, National Institute for Materials Science, Tsukuba 305-0044, Japan

[3]Research Center for Materials Nanoarchitectonics, National Institute for Materials Science, Tsukuba 305-0044, Japan

* To whom correspondence should be addressed, lee.jieun@snu.ac.kr.

**ABSTRACT: Atomic defects in solids offer a versatile basis to study and realize quantum phenomena and information science in various integrated systems. All-electrical pumping of single defects to create quantum light emission has been realized in several platforms including color centers in diamond and silicon carbide, which could lead to the circuit network of electrically triggered single-photon sources. However, a wide conduction channel which reduces the carrier injection per defect site has been a major obstacle. Here, we realize a device concept to construct electrically pumped single-photon emission using a van der Waals stacked structure with atomic plane precision. Defect-induced**

**tunneling currents across graphene and $NbSe_2$ electrodes sandwiching an atomically thin *h*-BN layer allow robust and persistent generation of non-classical light from *h*-BN. The collected emission photon energies range between 1.4 and 2.9 eV, revealing the electrical excitation of a variety of atomic defects. By analyzing the dipole axis of observed emitters, we further confirm that emitters are crystallographic defect structures of *h*-BN crystal. Our work facilitates implementing efficient and miniaturized single-photon devices in van der Waals platforms toward applications in quantum optoelectronics.**

Defects in solid-state systems provide a rich platform to study not only the fundamental properties of materials and their modifications but also the quantum nature of the isolated carriers, which find applications in quantum technologies.[1–5] Among various systems, two-dimensional materials provide an exceptional playground to study fundamental properties of defects regarding the dimensionality of the host material as well as their quantum nature for developing highly integrated quantum circuitries.[6–8] For example, defects in ultrathin hexagonal boron nitride (*h*-BN) have been shown to exhibit bright single-photon emission which can be used for quantum communication.[9,10] Also, single spins in *h*-BN defect sites have been found to probe minute physical quantities of proximate specimens in nanometer distance.[11–13] However, implementing quantum operation of *h*-BN defects in previous works has largely relied on the optical pumping source for the generation of carriers which imposes difficulties on the targeted and controlled manipulations of defects in integrated structures.

Electrical excitation of single defects, on the other hand, offers an appealing opportunity to selectively control quantum emitters in designated locations in electrical devices without the inclusion of the laser source that excites background signals.[14,15] Two-dimensional van der Waals materials provide a particularly suitable material platform for such device structure through the nanometer scale tailoring of the host material and electrodes.[16] In this structure, the direct carrier injection into the local defects using a voltage also leverages the charge control of defects[17,18] which will be potentially useful for synchronized and ultrafast manipulation of multiple quantum emitters with increased scalability. However, the electrical excitation of an isolated defect and single-photon generation in *h*-BN has not been demonstrated so far.

In this article, we report the direct electrical pumping of a single-photon emitter in *h*-BN embedded in an electronic device with atomically thin thickness. By fabricating a van der Waals heterostructure composed of *h*-BN, graphene, and $NbSe_2$, we established the electrical generation of single-photon emission by direct carrier injection. The electrically induced photon emission shows the intensity that scales linearly with the applied current and exhibits antibunching in the photon correlation measurement using the Hanbury Brown and Twiss interferometer. The asymmetric layer configuration of the van der Waals heterostructure formation also enables the stable generation of defect emission, allowing the analysis of characteristic photon energies and polarization axis of the emitters. Our work on the electrically pumped generation of *h*-BN single-photon emission will open possible applications of *h*-BN defects for miniaturized compact single-photon devices and chip-based quantum information science.

## RESULTS AND DISCUSSION

The device structure is illustrated in Fig. 1a. A few-layer $NbSe_2$ and graphene were used as top and bottom metallic electrodes, respectively, for applying tunneling current through optically active *h*-BN defect sites. To create optically active defects in *h*-BN, high-temperature annealing with $O_2$ gas flow[19] was performed on high-purity *h*-BN crystals.[20] A thin pristine *h*-BN spacer layer is placed between the optically active *h*-BN and top $NbSe_2$ as part of a sequential transfer of multilayer stacking. For the stacking method, geometry, and optical image of the heterostructure devices that we fabricated, please refer to Supporting Information. By applying a voltage across two electrodes, the charge current flow through the defect was generated which induces the electrical pumping of the emitter luminescence.

In the heterostructure, the combination of $NbSe_2$, a hole-like van der Waals metal,[21,22] and graphene is chosen to form an asymmetric electrical junction with *h*-BN in between, enabling the efficient injection of electron and hole carriers into the defect sites for the generation of light emission. In addition, the edges of two metallic electrodes were aligned with minimal overlap, confining the path of the tunneling current flow to go through the defects positioned within the aligned edges. Reducing the number of participating defects in the tunneling current generation further stabilizes the photon emission and allows the direct comparison of the emission intensity to the tunneling current.

With fabricated devices, we measured the electroluminescence (EL) from *h*-BN defect sites using an optical set-up equipped with a cryostat kept at 6.5 K while varying the applied voltage. Fig. 1b shows the electrically induced emission spectrum measured from one of the devices showing several narrow emission peaks brightened as a function of voltage. We name this device as Device 1. We note that Device 1 has the total thickness of *h*-BN corresponding to 7 nm between electrodes (Optically active *h*-BN: 5 nm, spacer *h*-BN: 2 nm). Above a threshold, the emission intensity of individual peaks gradually increases with the applied voltage. The spatial image measured by the charge-coupled device (CCD) imaging mode is shown in Fig. 1c. From the image depicting the reflection of a white light source, the boundaries of graphene and $NbSe_2$ electrodes can be identified and the localized EL is found at one corner of the stacked region.

Fig. 1d shows the line spectra of the emitters measured at the applied voltage of 0 and 28 V. The emission peaks around 1.5, 2.8, and 2.9 eV are identified as the zero-phonon lines (ZPLs) of single defects. The small peak around 2.6 eV is the phonon sideband of the ZPL peak at 2.8 eV, as confirmed from the energy separation (160 meV) matching with the longitudinal optical phonons of *h*-BN[23] and the spectral shift simultaneously observed from the two peaks in the time stability monitoring measurement (Figure S2, Supporting Information). On the other hand, the emitter with a sharp ZPL around 1.5 eV was free of a separate phonon sideband. The observed variation of phonon sidebands in *h*-BN could originate from the layered structure of the host material[24] and has been previously reported from optically excited *h*-BN emitters.[25,26] Regarding the line shape of the ZPL spectrum, all three emitters exhibited asymmetric ZPL with a higher spectral weight towards the low energy side which could have originated from low-energy acoustic phonon sidebands[24] or the voltage-induced peak broadening.

We then performed detailed measurements on the emitter with ZPL energy at 1.5 eV (named as E1). The high-resolution spectra of the emitter with varying voltage are presented in Fig. 2a and 2b, which show the peak emerging at the threshold voltage of about 26 V. By plotting the emission intensity and tunneling current together as a function of voltage (Fig. 2c), the intensity is found to scale linearly with the tunneling current, starting with the same threshold voltage. This suggests that tunneling current is induced by defect-mediated charge transport which directly contributes to the charge carrier supply at the emitter site. The linear relation between the tunneling current and emitted photon intensity is well observed from stable emitters such as emitter E1. However, even for a less stable emitter with intermittent blinking, the positive correlation between the tunneling current and emitted photon intensity could be observed (Figure S3), which shows that photon emission in our device is driven by the defect-induced current flow. Such a correlation between the emitter luminescence intensity and current is the direct evidence of the charge transport through a single defect, a distinct behavior enabled by the van der Waals device scheme which uses a thin host material.

By analyzing the evolution of the E1 emitter with voltage more closely, a Stark shift of the ZPL peak is observed as shown in Fig. 2a, representing the existence of the out-of-plane dipole moment of the emitter.[27–29] Also, the width of the ZPL emission increases as a function of the applied voltage (Figure S4 for more details). The peak broadening can be explained by the increase of the injection current which enhances the charge carriers in the vicinity of the defect site, resulting in the field fluctuation by surrounding mobile charges. This and the increased dipole moment of the emitter which couples more strongly to the field fluctuation[30,31] result in the inhomogeneous broadening of the ZPL emission. As shown in Figure S4, it is also found that the inhomogeneous broadening is more pronounced towards the low energy side for this emitter as voltage increases unlike optically excited $h$-BN emitters in this energy range.[32] However, we also note that the minimum linewidth broadening of emitters from our measurement was around 2 meV, still comparable to most optically excited $h$-BN emitters and electrically excited emitters in other solid-state systems.[33,34]

The single-photon emission property of the emitter E1 is confirmed by the second order correlation function measurement. Fig. 2d shows the measured correlation function with the exponential fitting result. After the background correction,[35] we obtained the $g^2(0)$ value of 0.25 ± 0.21, proving the quantum nature of the emitter pumped through an external voltage. The width of the dip extracted from the measured $g^2(\tau)$ curve was 18.2 ± 7.2 ns, longer than typical lifetimes obtained from optically excited quantum emitters in $h$-BN (~ a few ns).[7,36,37] This rather large width is because of the distinctive transition process associated with the electrical excitation. Since the current injection supplies carriers to the defect site across the bandgap of the material, it is highly probable that an additional charge-trapping level is involved during the excitation process, limiting the transition rate of the emitter. A more detailed explanation on the transition mechanism of the electrical excitation in comparison to the optical excitation is discussed in Figure S5.[14]

While recording the $g^2(\tau)$ curve of the emitter, the photon count rate measured at two detectors was ~ 700 cps at the measured current level of 6 nA. The total measurement time took about 25 hours which gives the coincidence count event number of ~ 75 at a large time delay. This photon emission count rate is smaller than that observed in electrically pumped emitters in bulk host materials[14,15] but comparable to emitters in other ultrathin crystals.[16,38] We assume that a rather high threshold voltage of Device 1 has prevented the increase of the applied voltage, resulting in a moderate injection current. For other devices we fabricated, we could observe emitters with lower thresholds (Table S1, Supporting Information) so it is highly probable that decreasing the threshold voltage would be possible by preparing cleaner interfaces between materials and improving the contact quality to increase the emitter intensity and scalability.[39,40]

Further, we checked the dipole axis direction of the emitter by the polarization-resolved measurement as shown in Fig. 2e. While applying a voltage, the polarization of the emitted light is measured through an analyzer in front of the CCD detector. By fitting the polar plot of the angle-resolved emission intensity with the function $I(\theta) = A\cos^2(\theta - \theta_0) + B$, where $\theta$ is the polarization angle of the emitted light measured relative to the high symmetry axis of the $h$-BN crystal, we find $\theta_0$ of about $60°$ for the emitter E1.

The emission wavelength and intensity of the emitter also showed good stability as a function of time which is shown in Fig. 2f. The emission intensity exhibits generally persistent values with a few intermittent brightening. Another emitter exhibiting better stability in terms of intensity is shown in supplementary Fig. S3 which is measured at a lower voltage (23 V). To understand the intermittent brightening of the emitter at a high voltage, we performed an analysis of the blinking statistics, by investigating the occurrence of the emission on and off states as a function of the duration time (Figure S6). From this measurement, we extracted that the effect of surrounding itinerant mobile charges is a dominant origin of the observed intensity fluctuation.[41]

In Supporting Information Section 6, we summarize the number of emitters that we observed from Device 1 and three additional devices. In total, more than 30 emitters were observed from Device 1 through the electrical pumping which exhibited good stability. In addition to the emitters shown in Fig. 1, some of the emitters were created after applying an extremely high voltage to the device above 60 V (Figure S7). The creation of new defect levels at this high voltage can have several possible origins. The charge state of the existing defects could have been changed due to a voltage-induced charge injection[17,18] or new defect sites could have been created by the breaking of B-N bonding in *h*-BN and coupling with adjacent layers[42–44] or ions.[45–47] This suggests that the generation of local electrical pulse at the device level could be applied to controllably activate an emission center for the deterministic emitter creation.[48,49]

In Fig. 3, we present graphs showing the emission energy of the observed emitters arising at positive (red) and negative (blue) applied voltages for Device 1 to understand the charge supply mechanism of the heterostructure. From the graphs, two evident characteristics can be found. First, the number of emitters appearing at positive voltages far exceeds that appearing at negative voltages. Second, at high photon energies, most emitters are observed under the application of positive voltages with only a few exceptions.

Our observation is fully compatible with the band diagram of the device structure shown in Fig. 3a. Because of the different work functions of graphene ($W_G = 4.5$ eV)[50,51] and $NbSe_2$ ($W_{NbSe_2} = 5.9$ eV),[52] the heterostructure at zero bias creates a built-in band tilting in *h*-BN and the Fermi level of $NbSe_2$ at this configuration becomes closer to the *h*-BN valence band maximum. By applying a finite voltage, current flow supplies electrons and holes into the defect sites, leading to photon emission. This process occurs more efficiently under a positive bias because $NbSe_2$, a hole metal, preferentially provides hole carriers for the tunneling current. The supply of holes from $NbSe_2$ and electrons from graphene allows the stable generation of photons at the defect location. At a negative voltage, electron carriers are less supplied, resulting in a lower probability of photon emission, which is more pronounced for emitters with higher transition energies.

With information on the wavelength and polarization axis of each emitter, we could further categorize the emitters into three different groups, which could have distinct crystallographic structures (Fig. 4). These three groups are the long-wavelength group (Group 1) with the energy 1.4 – 1.7 eV, the middle-wavelength group (Group 2) with the energy 1.9 – 2.4 eV, and the short-wavelength group (Group 3) with the energy 2.4 – 3.0 eV. Different groups show characteristic photon energy, the shape of phonon sidebands, and polarization dependence, implying that they have different origins.

Group 1 emitters showed sharp ZPL peaks with small phonon sidebands and randomized distribution of the polarization axis. The emission energy range and the shape of the spectrum match well with that of oxygen-related defect centers.[32,53] The abundance of group 1 emitters in our experiment could be the result of the oxygen-flow annealing during the fabrication process. From other devices fabricated using argon-annealed *h*-BN layers, strikingly contrasting emitter distribution is found (Figure S8). Our work thus highlights the accessibility of an EL experimental tool to probe oxygen-related defects in *h*-BN crystal. On the contrary, group 3 emitters showed phonon sidebands about 160 meV apart from the ZPL and axes of the linear polarization which are generally separated by a 60-degree interval. The polarization axis is particularly more concentrated around 120° which could be the result of the uniaxial strain frequently found in layered materials.[54] The observed polarization dependence agrees with the recently reported emitters induced by e-beam irradiation[55,56] or carbon ion implantation.[57] These suggest that group 3 emitters have a high possibility to originate from carbon-related color centers. Group 2 emitters, in between, showed a wide range of variation in both the spectral shape and polarization distribution. Further research on electrically driven defect emission created by more controlled fabrication methods will unveil additional information on the structural identification, carrier transport, and emission mechanism of *h*-BN defect centers.

## CONCLUSIONS

In conclusion, we demonstrated all-electrical generation of single-photon emission from *h*-BN by incorporating a van der Waals integrated device scheme. The combination of graphene and $NbSe_2$ as electrode materials in an asymmetric device structure enables the stable generation of photon emission in a wide range of spectral windows, allowing studies on various types of emitters with electrical excitation. We anticipate expanding the range of two-dimensional materials in heterostructure fabrication design involving *h*-BN defects will promote the functionality and applicability of van der Waals quantum optoelectronics for advanced quantum technologies. Through integrable single-photon sources in ultrathin materials with electrical excitation capabilities, further progress in electrically driven quantum communication[58] and spin defect operations[59] will be possible.

## METHODS

All optical measurements are performed with a home-built confocal microscopy set-up. The luminescence signal from the sample is collected using an objective lens with the numerical aperture of 0.65, passed through the collection path, and entered either a spectrometer or a Hanbury Brown and Twiss interferometer set-up.

For the spatial characterization of the emitters, the image of the electrically excited emission is collected by setting the grating of the spectrometer in the zeroth-order position, which functions as a mirror to detect the 2D map of the emitters at the CCD position. Outlines of the sample boundaries are identified by the reflection image using a white-light source. To detect the voltage-dependent emitter spectra, photons emitted from the defect centers are dispersed by a diffraction grating in the spectrometer and recorded by the CCD in the spectrum mode. The polarization dependence of the emitted signal is obtained by using a polarizer and a half-wave plate in the rotating mount in the collection path.

In the Hanbury Brown and Twiss interferometer set-up, the emission signal is split by a 50:50 beam splitter in the collection path and recorded by two avalanche photodiodes (APDs). The APDs are connected to a time-correlated single-photon counting device for monitoring the second order correlation function and temporal stability of the emitter luminescence.

## ASSOCIATED CONTENT

**Supporting Information**

Device fabrication; time-dependent spectral shift of emitters in Fig. 1; real-time measurement of tunneling current and defect emission; spectral shape and linewidth evolution as a function of voltage; transition mechanism of electrically and optically excited emitters; summary of $NbSe_2$/*h*-BN/Graphene stacked devices; blinking statistics of an emitter; emitters created after applying a high voltage; comparison between $O_2$-annealed and Ar-annealed *h*-BN emitters;

## ACKNOWLEDGEMENTS

We acknowledge support from the National Research Foundation of Korea (NRF) grant funded by the Korea government (MSIT) (No. 2020R1A2C201133414, No. 2021R1A5A103299614, No. RS-2024-00356893, and No. RS-2024-00413957) and the Institute for Basic Science (IBS) in Korea (No. IBS-R009_D1). The research is further supported by the ITRC (Information Technology Research Center) support program (RS-2022-00164799) supervised by the IITP (Institute for Information & communications Technology Promotion) of Korea. K.W. and T.T. acknowledge support from the JSPS KAKENHI (Grant Numbers 21H05233 and 23H02052) and World Premier International Research Center Initiative (WPI), MEXT, Japan.

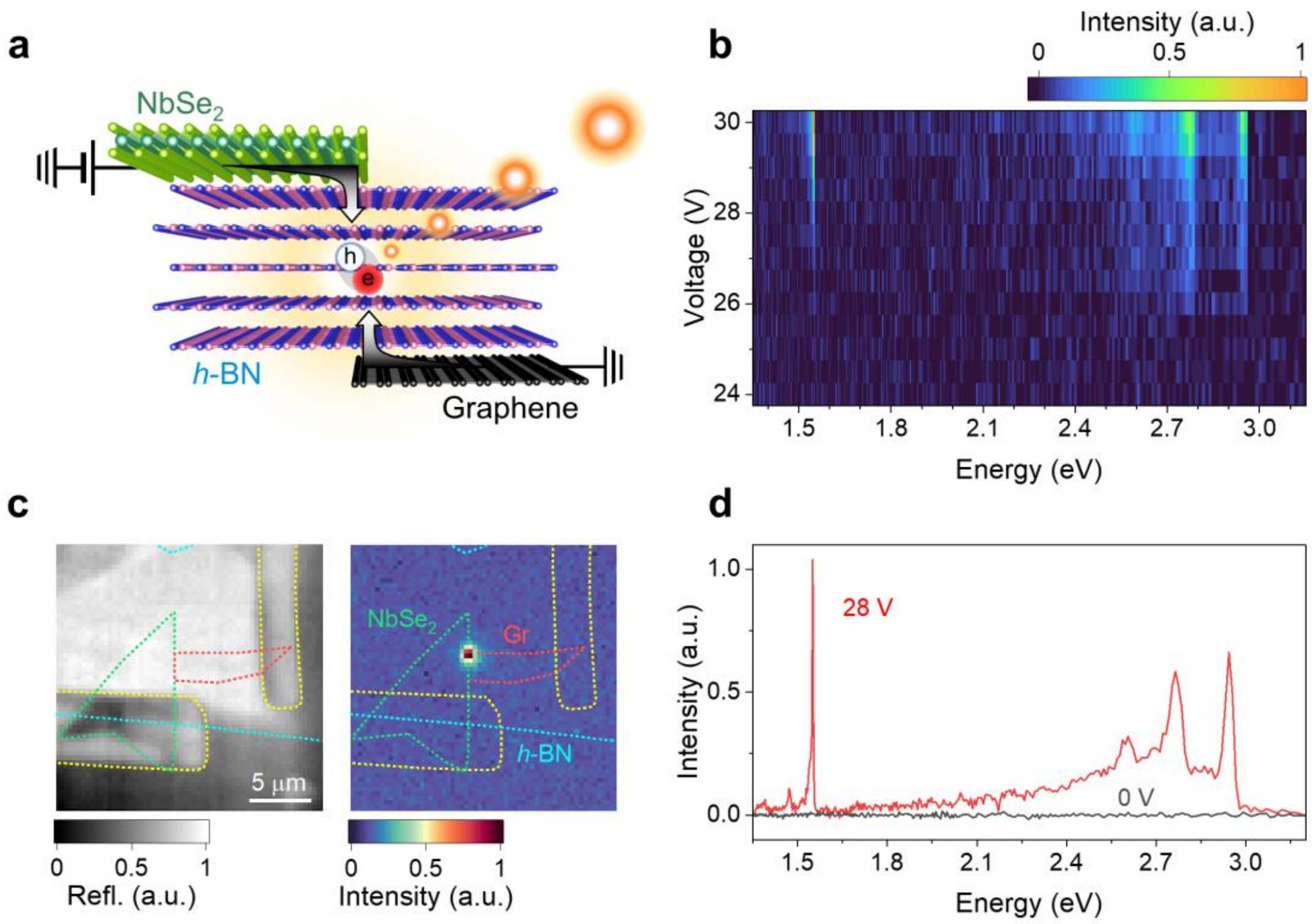

**Figure 1. Electrically pumped emitters from *h*-BN heterostructure device. (a)** Schematics of the *h*-BN heterostructure device with $NbSe_2$ and graphene electrode. Red and white filled circles represent electron and hole carrier at the defect site, respectively. Orange open circles indicate the single-photon generation. **(b)** Color map of the voltage-dependent spectra. All data are normalized by the maximum intensity. **(c)** 2D spatial image obtained by a CCD from white light reflection of the device (left) and electroluminescence emission at 28 V (right). Blue, green, red, and yellow lines are optically active *h*-BN, $NbSe_2$, graphene and Au electrodes, respectively. **(d)** Emission spectra measured at 0 (black) and 28 V (red).

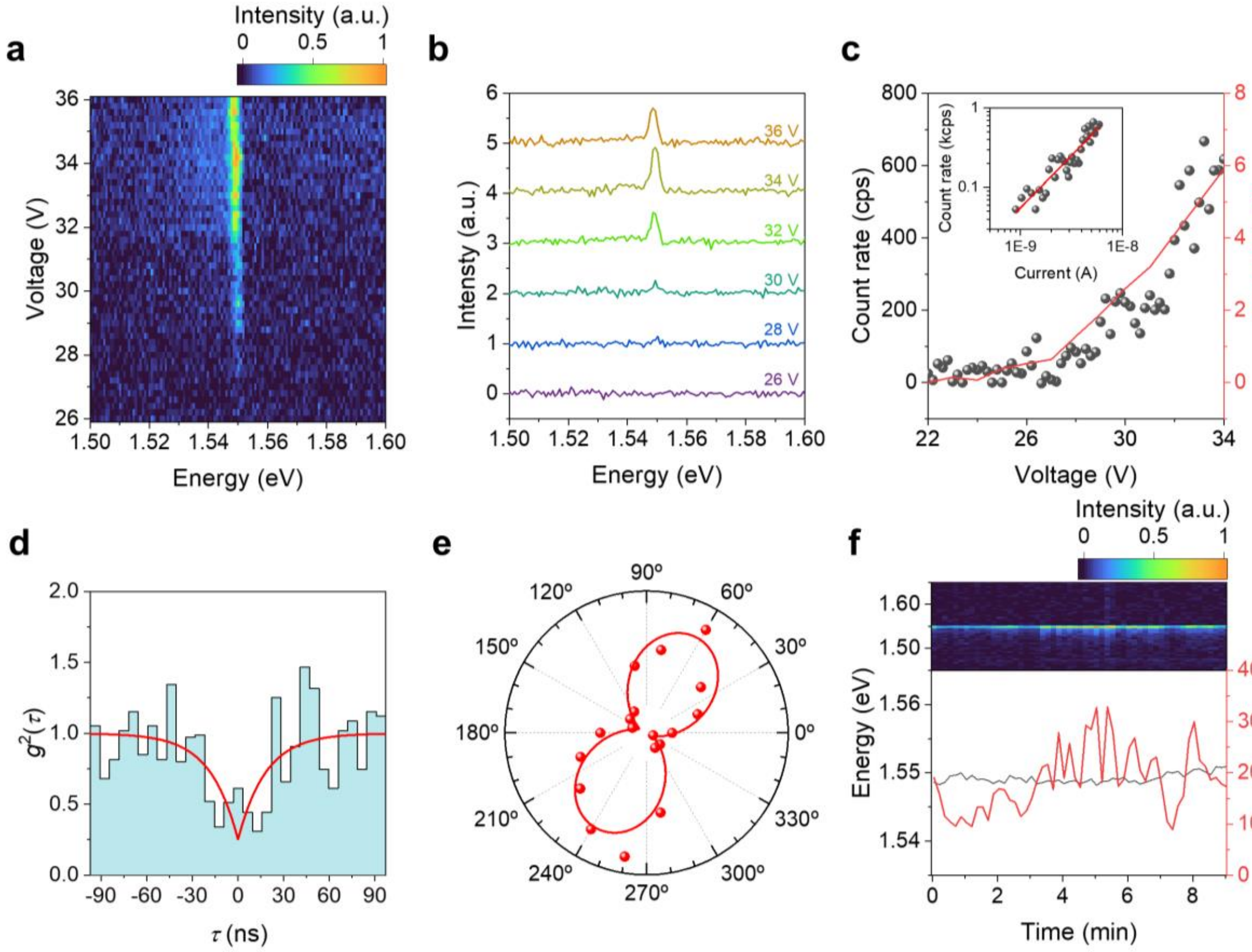

**Figure 2. Optical and electrical characteristics of the emitter E1. (a)** High resolution color map of the voltage-dependent spectra. **(b)** Spectra of the emitter at selected voltages. **(c)** Emitter count rate (black dots) and channel current (red line) as a function of voltage. Inset: Log scale plot of intensity vs current. The slope of the linear fit is about 1.31. **(d)** Second order correlation function $g^2(\tau)$ data under the injection current of 6 nA (filled area) and the exponential fit function (red line). **(e)** Polar plot of the emission intensity (red dots) and the fit result using $\cos^2(\theta)$ function (red line). **(f)** Time stability measurement spectra of the emitter at 30 V (upper) and corresponding center energy and intensity as a function of time.

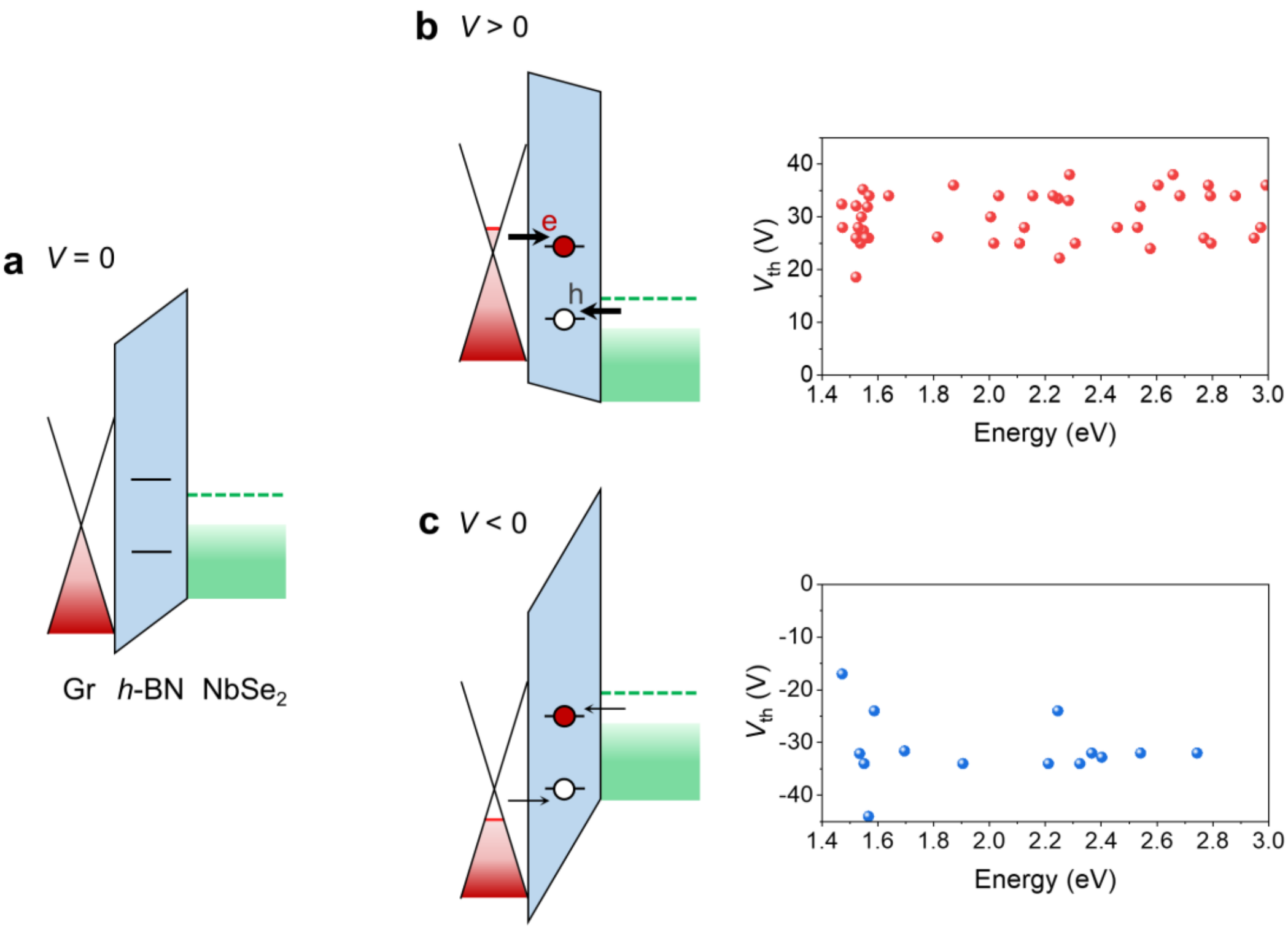

**Figure 3. Charge injection mechanism and emission energy distribution. (a)** Band diagram of graphene, *h*-BN, and $NbSe_2$ stacked heterostructure at zero applied voltage. A Red solid line is the Fermi level of graphene and a green dashed line is the energy maximum of the partly filled hole band of $NbSe_2$. The built-in band tilting at zero voltage reflects the work function difference between graphene (4.5 eV) and $NbSe_2$ (5.9 eV). **(b)** Band diagram under the positive voltage (left) and the distribution of emitters as functions of threshold voltage and emission energy (right). Injected electron and hole carriers are indicated by red and white filled circles, respectively. **(c)** Band diagram under the negative voltage (left) and the distribution of emitters as functions of threshold voltage and emission energy (right).

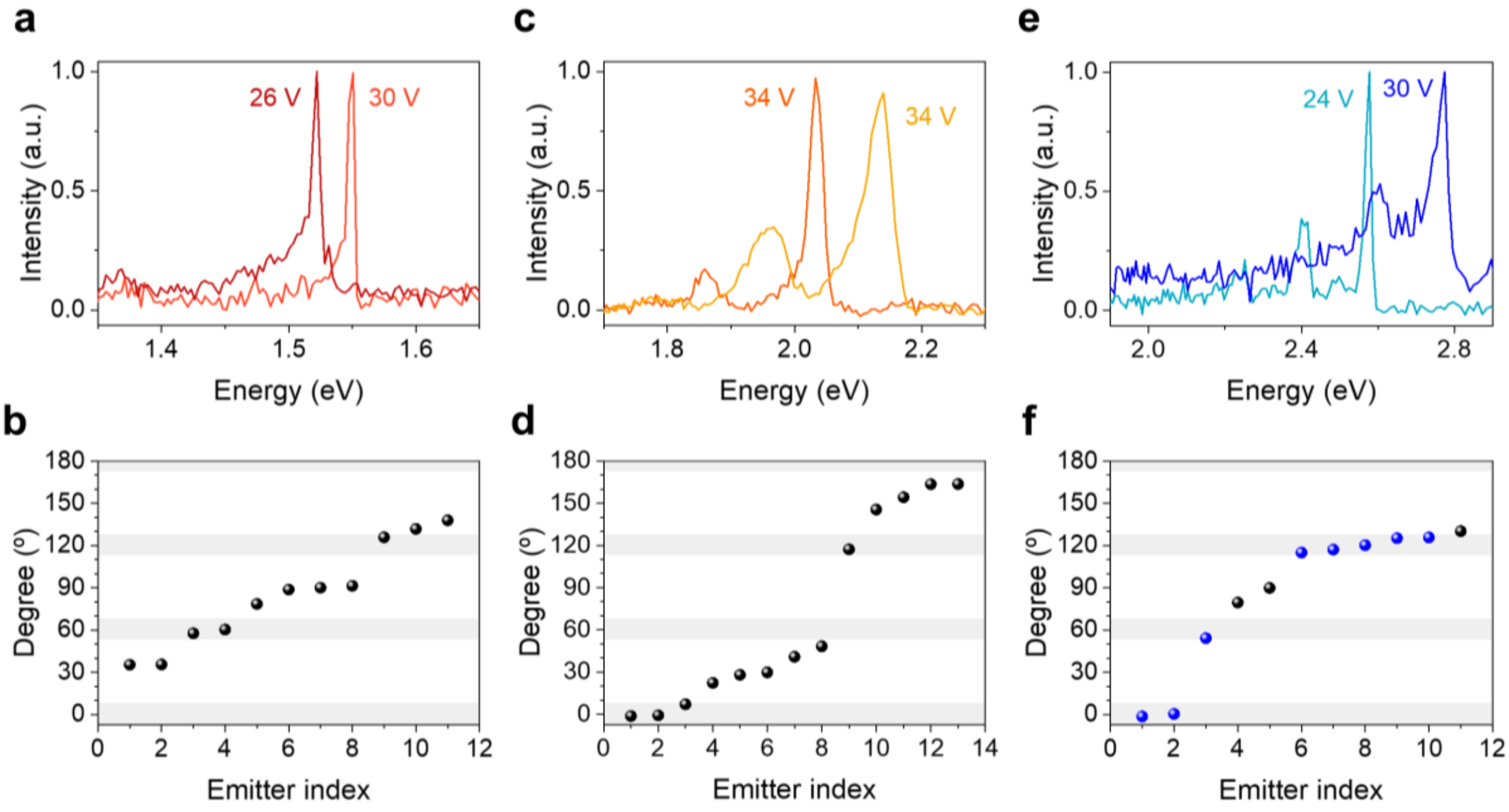

**Figure 4. Representative emitter spectra and distribution of emission polarization.** Representative spectra and dipole moment axes distribution of emitters from three different groups: **(a,b)** Group 1: emitters distributed in 1.4 – 1.7 eV, **(c,d)** Group 2: emitters distributed in 1.9 – 2.4 eV, **(e,f)** Group 3: emitters distributed in 2.4 – 3.0 eV. The representative emitter spectra shown in (c) were collected at the same applied voltage but at different times and voltage scans. In the polarization distribution graphs, 0° corresponds to a high symmetry axis of the *h*-BN crystal identified from an optical microscope image. Gray transparent backgrounds are shown as a guide to eyes for high symmetry axes of the *h*-BN crystal. Blue circles indicate emitters with dipole orientations aligned with the crystal axes.

## TABLE OF CONTENTS

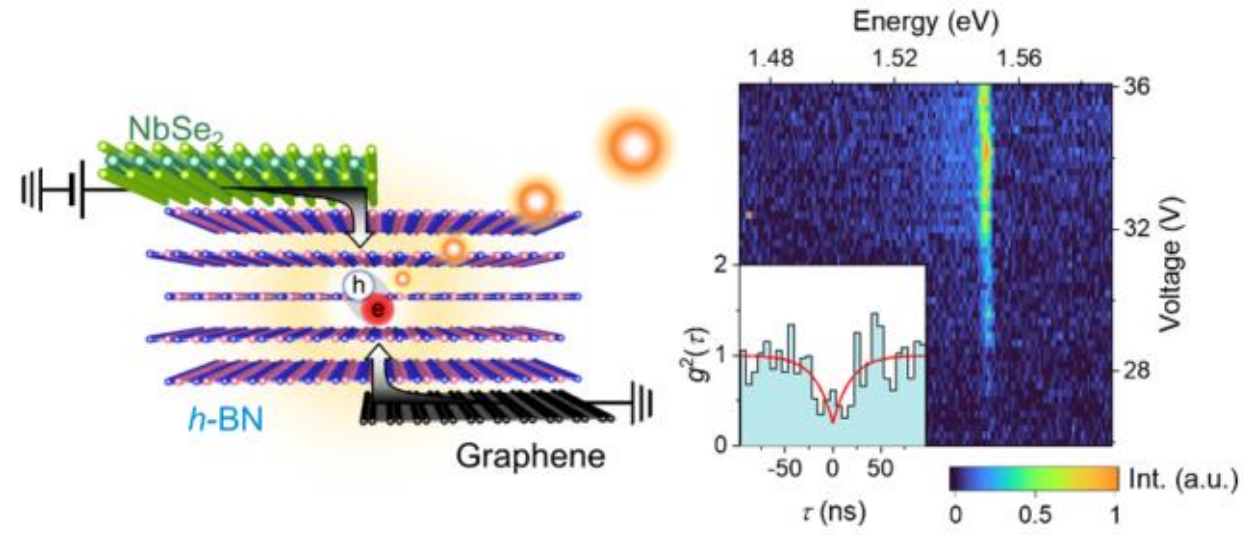